\documentclass[10pt,showpacs,twocolumn,
preprintnumbers,amsmath,amssymb,
aps,prd,nofootinbib,eqsecnum,a4paper]{revtex4}
\usepackage{epsfig}
\usepackage{graphicx,epsf}
\usepackage{color}
\usepackage{bm}
\def\be{\begin{equation}}
\def\ee{\end{equation}}
\def\bea{\begin{eqnarray}}
\def\eea{\end{eqnarray}}
\def\bi{\begin{itemize}}
\def\ei{\end{itemize}}
\def\p{\partial}

\def\L{{\pounds}}

\def\Q{{{\cal{Q}}}}
\def\cs2{c_{\rm{s}}^2}
\def\wt{\widetilde}
\newcommand\eq[1]{Eq.~(\ref{#1})}
\newcommand\eqs[1]{Eqs.~(\ref{#1})}
\def\com{{\mathrm{com}}}
\def\tc{{t_\mathrm{c}}}

\def \beg {\begin{enumerate}}
\def \en {\end{enumerate}}

\def\M0{{\cal M}_0}
\def\Me {{\cal M}_{\epsilon}}

\begin{document}

\title{Comments on gauge-invariance in cosmology}

\author{Karim A. Malik$^1$}
\author{David R.~Matravers$^2$}

%
\affiliation{
$^1$Astronomy Unit, School of Physics and Astronomy, Queen Mary University
of London, Mile End Road, London, E1 4NS, United Kingdom\\
$^2$Institute of Cosmology and Gravitation, University of Portsmouth,
Dennis Sciama Building, Portsmouth, PO1 3FX, United Kingdom}
\date{\today}

\begin{abstract}
We revisit the gauge issue in cosmological perturbation theory, and
highlight its relation to the notion of covariance in general
relativity. We also discuss the similarities and differences of the
covariant approach in perturbation theory to the Bardeen or metric
approach in a non-technical fashion.
\end{abstract}

\pacs{98.80.Jk, 98.80.Cq \hfill  arXiv:1206.1478v4}

\maketitle


\section{Introduction}
\label{intro}

Perturbation theory is these days a standard tool in theoretical
cosmology. Since the introduction of the theory more than sixty years
ago, the ``gauge issue'' has been troubling cosmologists.
Although the problem has been resolved by now, at all orders in
perturbation theory, there still seems to be some confusion as to its
origins. We aim at clarifying these issues in this short note, by
highlighting previous results that might be well known to the
specialist, but not necessarily to the wider community.  \\

After the pioneering work by Lifshitz in 1946 \cite{Lifshitz1946},
Bardeen in 1980 \cite{Bardeen80} demonstrated how the gauge issue can
be rigorously solved in the metric based approach. The reviews by
Kodama and Sasaki \cite{KS} and Mukhanov, Feldman and Brandenberger
\cite{MFB} further contributed to the success of this approach.
More recently the gauge issue has also been studied beyond linear
order in cosmological perturbation theory, leading to a rich body of
work using the Bardeen and related gauge-invariant formalisms
\cite{Bruni96,Mukhanov96,MW2003,Noh2004,Bartolo_review,Nakamura2007,MM2008,MW2008,CMMN2011}.

A different approach has been developed, which is usually referred to
as the ``covariant'' approach, following Ellis and Bruni
\cite{Ellis:1989jt} (and earlier work by
\cite{sachs,hawking66}). However, as we show below, also the covariant
approach has to face the gauge issue if a ``background'' spacetime is
introduced.\\

But what do we mean by gauge dependence in cosmological perturbation
theory, where does it actually come from, and what makes it so
difficult?
In cosmological perturbation theory we split quantities into a
background and small perturbations, both for the matter variables and
the metric. Gauge dependence then stems from requiring a unique
background spacetime on which the background quantities are defined,
that does not follow the coordinate or gauge transformation, whereas
the perturbations do obey the transformation. This allows us to talk
about e.g.~\emph{the} Friedmann-Robertson-Walker (FRW) background
around which we perturb our quantities.

Before we define once more what we mean by gauge-invariance in the
setting of cosmological perturbation theory below, let us briefly
state what it is not. It should not be mistaken for or mixed up with
{covariance}, the requirement that the governing equations do not
depend on the coordinate system chosen. Indeed, it is well known that
covariance can be broken in cosmology without conceptual or other problems
arising, e.g.~in order to choose a particular coordinate system that
makes the calculation simpler \cite{EM1995}.

Gauge-invariance in cosmological perturbation theory should also not
be confused with the gauge choice in standard General Relativity
(GR). We can choose four arbitrary coordinate functions in the metric
tensor (thanks to the four constraint equations in Einstein's field
equations, see e.g.~Ref.~\cite{MTW}). 
In cosmological perturbation theory the symmetries of the physical
spacetime, namely homogeneity and isotropy, lead us to choose a
background that is FRW (with zero background curvature for
convenience), and we can then use covariance to choose our coordinates
in the background such that the line element has a particularly simple
form, i.e.
\be
ds^2=dt^2+a^2(t)\delta_{ij}dx^idx^j\,,
\ee
where $a$ is the scale factor and $\delta_{ij}$ the flat background
metric.
This is related to the ``geodesic slicing'' in numerical
relativity, which is selected by choosing a lapse function normalised
to unity, and choosing a vanishing shift vector selects Gaussian
normal coordinates
(see e.g.~Ref.~\cite{Baumgarte2009}).
Hence the ``standard GR coordinate freedom'' is used, actually used
up, in choosing and specifying the background. \\

This note is aimed at the non-specialist and we try to keep our
discussions as non-technical as possible, working rather by way of
example than by introducing theorems. Nevertheless, in the following
two sections we have to define gauge-dependence and gauge-invariance
in a more rigorous way than above.  In Section \ref{relating} we show
how quantities in the covariant formalism are related to the ones in
standard metric perturbation theory, and comment briefly on the issue
of non-locality in perturbation theory in Section \ref{nonlocal}, and
conclude with a short discussion in Section \ref{discussion}. Finally
we highlight in Appendix \ref{example} the difference between
covariance and gauge-invariance.

We use predominantly coordinate time $t$, and denote derivatives with
respect to $t$ with a dot. Conformal time $\eta$ is related to $t$ by
$dt=a\ d\eta$, where $a$ is the scale factor. Greek indices $\mu,
\nu$, and $\lambda$ range from $0$ to $3$, lower case Latin indices,
$i$, $j$, and $k$, have the range $1,2,3$.

\section{The origin of gauge dependency in perturbation theory}
\label{origin}

So far we have mentioned that there is a problem in cosmological
perturbation theory often referred to as the ``gauge issue'', but in
the previous section we only sketched it very roughly.
Let us therefore now turn to the origin of gauge dependence and define
what we mean by \emph{gauge}, \emph{gauge dependence}, and \emph{gauge
  invariance}. We shall use the FRW background universe as an example
spacetime. As stated above, we shall attempt to be as non-technical as
possible in this note, and refer the interested reader to the seminal
papers by Stewart and Walker \cite{stewartwalker} and Stewart
\cite{stewart} for technical details, and to Ref.~\cite{MM2008} whose
notation we follow where possible.\\

The physical domain (``the universe'') in which we work in cosmology can
be regarded as a single manifold $\Me$.  In perturbation theory,
following Ref.~\cite{stewartwalker} we introduce another, fiducial,
manifold $\M0$ (frequently referred to as the ``background''
manifold). A correspondence or map (one-to-one mapping or
diffeomorphism) between $\Me$ and $\M0$ is referred to as a {\it
gauge} and the correspondence is generated by a vector field, the
\emph{gauge generator}.

\begin{figure}
\begin{center}
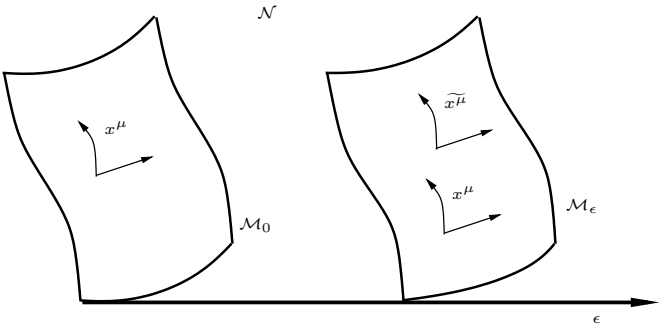
\caption[pix]{\label{fig1} 
The manifolds $\M0$ and $\Me$, embedded in
  the higher dimensional manifold $\cal N$. The small parameter
  $\epsilon$ is an additional dimension (or label) in $\cal N$. 
}
\end{center}
\end{figure}

As illustrated in Fig.~\ref{fig1} the manifolds $\Me$ and $\M0$ are
treated as being embedded in the higher dimensional manifold
$\mathcal{N} = \mathcal{M} \times \mathcal{R}$.  The parameter
$\epsilon$ allows us to distinguish between the manifolds $\M0$ and
$\Me$, the former, the background, being at label $\epsilon = 0$.  We
can also introduce coordinate systems. We start out with $ \{x^{\mu}\}$ on
$\M0$, and identifying the points on $\Me$ with those on $\M0$ we can
also label the points on $\Me$ with the coordinates $\{x^{\mu}\}$. Then
relabelling the points on $\Me$ induces a \emph{gauge transformation}.

The gauge issue arises from splitting quantities, defined on $\mathcal
{N}$ into a background part, on $\M0$, which usually depends on fewer
dimensions (or coordinates) than the physical part (on $\Me$) which
defines the perturbation which usually depends on the full set of
dimensions (coordinates) of the spacetime. When we change the
coordinate system on the physical space $\Me$ by a small amount
$x^{\mu} \rightarrow \tilde{x^{\mu}}$ and require that the coordinates
of the background remain unperturbed, to keep a unique background
spacetime (e.g.~the FRW background in standard cosmology) the
perturbed quantities on $\Me$ will undergo a gauge transformation
generated by this relabelling of coordinates. \\

Let us rephrase the above in a slightly more technical form. We can split
any tensorial quantity $\Q$ into a background part, $\Q_0$ and a small
perturbation $\delta\Q$, that is
\be
\label{Qsplit}
\Q\equiv \Q_0+\delta\Q\,.
\ee
Note that the quantity $\Q$ is defined on physical manifold $\Me$
(``background+ perturbation''), whereas $\Q_0$ is defined on $\M0$
(``background'', with $\epsilon=0$).

We can now change the coordinate system $x^\mu$ on $\Me$ to a new
system $\tilde x^\mu$ which is related to the old one by a small
amount, $\delta x^\mu$ that is
\be
\label{coord_change}
\tilde x^\mu = x^\mu +\delta x^\mu\,.
\ee
This change will also induce\footnote{We are choosing here the
  ``passive''point of view, to calculate the change of the variables
  under a gauge transformation. In this approach, one calculates the
  effect that a change of coordinate system, Eq.~(\ref{coord_change}),
  has on the variables. In the active approach the change in the
  variables is calculated directly from the ``action'' of the gauge
  generator $\delta x^\mu$ (see Ref.~\cite{MM2008} for details).} (in
  general) a change in the perturbation $\delta\Q$, that is
\be
\label{deltaQchange}
\delta\Q(x)\to\wt{\delta\Q}(\tilde x)\,,
\ee
however, the background quantity remains unchanged,
\be
\label{Qback}
\Q_0(x)=\wt{\Q_0}(\tilde x)\,.
\ee
Studying the transformation behaviour of the perturbed quantities
allows us to ``fix'' or correct this different and disparate behaviour
of the quantities (i.e.~the background quantities remain fixed, the
perturbations are allowed to vary), and construct gauge-invariant
quantities (free of gauge-artefacts, the $\delta x^\mu$ defined in
\eq{coord_change}). Note that $O(\delta\Q)=O(\delta
x^\mu)=O(\epsilon)$.\\

The terms ``gauge'' and ``gauge invariance'' are used in different
ways in the literature, we mentioned the use of the term in GR in the
previous section. Another definition of gauge invariance, due to
Stewart and Walker, is sometimes also referred to as
\emph{identification gauge invariance} (``i.g.i.''~in Stewart and
Walker \cite{stewartwalker}) or ``strong'' gauge invariance. For a
perturbed quantity to be identification gauge invariant its background
part has to be either, vanishing\footnote{Either the background part
has to be zero, or its Lie derivative with respect to the
gauge-generator has to be zero, i.e.~$\L_{\delta x^\mu}\Q_0=0$, to be
precise \cite{stewartwalker}. The restrictive notion of i.g.i.~can
be extended to order $n$ in perturbation theory by requiring that
$\L_{\delta x^\mu}\Q_{m}=0$, where $m$ is in the range of $0\ldots
n-1$ \cite{Bruni96}. Note that the Lie derivative has to act on
the whole tensorial quantity, and not just a particular component of
it.}, or constant, or constructed out of linear combination of
Kronecker deltas \cite{stewartwalker}. This is usually referred to as
the ``Stewart-Walker Lemma''. This Lemma is rather restrictive, it
doesn't allow for any gauge-artefacts (the gauge generators $\delta
x^\mu$) to appear in the transformation equations for the perturbed
variables.

The definition of gauge invariance we use here, following Bardeen, is
weaker than the identification gauge invariance: we require that there
are no gauge artefacts (the ``gauge generators'', or $\delta x^\mu$
defined in \eq{coord_change}) left in the governing equations, if they
are subjected to the transformation (\ref{coord_change}). Similarly, a
gauge-independent quantity is one that doesn't ``pick up'' gauge
artefacts if subjected to the transformation \eq{coord_change}.  Hence
quantities can be gauge-invariant, in this more general sense, but not
satisfy the Stewart-walker lemma. We will illustrate this with an example
in the following section.

\section{Metric or coordinate based perturbation theory}
\label{standard}

In Section \ref{origin} we discussed the behaviour of tensorial
quantities under a gauge transformation in very general terms. We now
derive how perturbations transform in a particular spacetime, using
the transformation behaviour of the first order or linear density
perturbation in a FRW background, where the background momentum
density vanishes, as an example\footnote{We use the energy density
here as an example for the transformation behaviour of a scalar
quantity. In more complex spacetimes than FRW, the energy density has
to be derived from the energy-momentum tensor specifically.}, and then
discuss how to construct gauge-invariant quantities. The derivation
follows as closely as possible Ref.~\cite{MM2008}. \\

Under the transformation \eq{coord_change} we get, using the passive
approach, that the energy density in the new (``tilde'') system is
related to the density in the old system as
\bea
\tilde\rho\left(\tilde x^\mu \right)
&=&\tilde\rho\left(x^\mu +\delta x^\mu\right)\nonumber\\
&=&\tilde\rho\left(x^\mu\right)+\tilde\rho_{,\mu}\delta x^\mu\,,
\eea
where we used a Taylor expansion, truncating here and in the following
at linear order, in the second line. Then using the expansion
\eq{Qsplit} for the energy density, and also applying \eq{Qback} to
the energy density, that is we require $\tilde\rho_0(t)=\rho_0(t)$
(the FRW background is time dependent only) we arrive at the
standard result
\be
\label{gaugetrans_rho}
\wt{\delta\rho}={\delta\rho}-\dot\rho_0 \delta t\,,
\ee
the transformation of the energy density under a gauge
transformation, at linear order. Here we used the standard $3+1$ split
for the gauge generator $\delta x^\mu$, that is $\delta x^\mu\equiv
[\delta t, \delta x^i]$.

From the transformation of $\rho$, \eq{gaugetrans_rho}, we can also
see how the Stewart-Walker lemma ``works''. If the quantity in the
background is zero or constant, the second term in \eq{gaugetrans_rho}
would also be zero, and hence the perturbation in the new and the old
coordinate system would be identical, that is gauge-invariant in the
strong sense.\\

The perturbed FRW metric for scalar perturbations and a flat
background is given by
\bea
\label{pert_metric}
ds^2&=&-(1+2\phi)dt^2+2aB_{,i}dtdx^i\nonumber\\
&&+a^2\big[(1-2\psi)\delta_{ij}+E_{,ij}
\big]dx^idx^j\,.
\eea
Here $\phi$ is the lapse function, $B$ and $E$ describe the shear, and
$\psi$ is the curvature perturbation related to the perturbed
intrinsic curvature of spatial 3-hypersurfaces by ${}^{(3)}R=4\nabla^2
\psi/a^2$ (see e.g.~Ref.~\cite{MW2008}).

To keep the algebra in this note at a minimum we do not show here how
to derive and calculate the transformation behaviour of the metric
tensor and refer the interested reader to Refs.~\cite{MM2008} and
\cite{MW2008} for details. We simply quote the result for e.g.~the
curvature perturbation here,
\be
\label{gaugetrans_psi}
\wt \psi=\psi +H \delta t\,,
\ee
where $H\equiv\frac{\dot a}{a}$.  \\

We now have two perturbed quantities and their transformations at our
disposal, and this allows us to construct a gauge-invariant quantities
as follows.
Equations \ref{gaugetrans_rho} and \ref{gaugetrans_psi} let us
construct two related gauge-invariant quantities, since we can decide
to choose two particular gauges or hypersurfaces: working on uniform
density hypersurfaces, defined as $\wt{\delta\rho}=0$, we get from
\eq{gaugetrans_rho} for the temporal gauge shift $\delta t$,
\be
\label{delta_t_rho}
\delta t\Big|_{\wt{\delta\rho}=0}=\frac{\delta\rho}{\dot\rho_0}\,.
\ee
Substituting this expression into \eq{gaugetrans_psi} we get 
\be
\label{def_psi_rho}
\psi\Big|_{\wt{\delta\rho}=0}
=\psi +\frac{H}{\dot\rho_0}\delta\rho\,,
\ee
the curvature perturbation on uniform density hypersurfaces.

It can be shown, that this quantity is conserved on very large scales
for adiabatic systems that is with barotropic equation of state
(independent of the underlying theory of gravity) \cite{WMLL}, and is
usually denoted by $\zeta$, where for historic reasons
$-\zeta=\psi\Big|_{\wt{\delta\rho}=0}$.

Similarly, instead of working on uniform density hypersurfaces, we can
choose uniform curvature hypersurfaces, defined as
$\wt{\psi}=0$, which gives from \eq{gaugetrans_psi}
\be
\label{delta_t_psi}
\delta t\Big|_{\wt{\psi}=0}=-\frac{\psi}{H}\,.
\ee
Substituting this expression into \eq{gaugetrans_rho} we get 
\be
\label{def_rho_psi}
\delta\rho\Big|_{\wt{\psi}=0}
=\delta\rho +\frac{\dot\rho_0}{H}\psi\,,
\ee
the density perturbation on uniform curvature hypersurfaces.

Both quantities defined in \eqs{def_psi_rho} and \ref{def_rho_psi}
above are gauge-invariant, that is invariant under a transformation of
the form \eq{coord_change}. This can be readily seen from the
respective definitions, as in both cases the gauge-artefacts $\delta
t$, that the ``raw'' quantities would pick up, cancel in the
respective combinations.
It is this definition of gauge-invariance, namely that the
gauge-invariant quantities do not pick up gauge-artefacts $\delta
x^\mu$, that we use and which is, arguably, more popular these days
\footnote{Note that neither the ``raw'' quantities nor their
  gauge-invariant combinations satisfy the Stewart-Walker lemma: the
  energy density has in general a non-zero background part, and the
  curvature perturbation is related to the expansion and hence the Hubble
  parameter in the background.}.
This notion of gauge-invariance can also be readily extended to higher
order (see e.g.~Refs.~\cite{MM2008,MW2008} for examples at second
order, and Ref.~\cite{CM2009} for examples at third order).

\section{The covariant approach to perturbation theory}
\label{covariant}

We now turn to the``covariant approach'' to cosmological perturbation
theory and how it deals with any potentially ensuing gauge issues.  One
might be tempted, given the approach's name, that here the gauge issue
is avoided altogether, after all GR doesn't suffer from the
perturbation theory gauge issue besides the ``normal'' coordinate
freedom, as pointed out in Section \ref{intro}.

The covariant approach to perturbation theory
\cite{sachs,hawking66,Ellis:1989jt,Bruni:1992dg,Challinor:1998aa,LV2006,Tsagas:2007yx} allows in
principle to avoid the gauge issue, as it does not require to specify
the metric from the beginning of the calculation.
However, as soon as one imposes a split into a particular background
and perturbations, the same issues as described above in Section
\ref{origin} and troubling the metric approach will ``haunt'' the
covariant approach.

One possibility to remove gauge artefacts is to use the Stewart-Walker
lemma, and use quantities that vanish in the background. Hence in FRW
we can construct the spatial gradient of a scalar, and since the
background is time dependent only, this has no correspondence in the
background (i.e.~removes gauge terms, see below).
However, this only works at linear order in a standard FRW background:
if we think perturbatively again, as the second order quantities
``live'' in the first order time and space dependent background, which
does allow gradients. This is also the reason why second order tensor
perturbations are no longer gauge-invariant, as they have first order
tensor perturbations acting as a background\cite{MM2008}.

\subsection{Relating ``covariant'' and ``Bardeen'' quantities:}
\label{relating}

We can now relate quantities in the covariant formalism and in the
Bardeen formalism (see e.g.~Refs.~\cite{Bruni:1992dg}
and \cite{LV2006} for earlier work on this topic).
In the covariant formalism quantities (gradients) are projected down
onto spatial three sections using the projection tensor defined below,
\eq{project}, relative to some vector field, usually chosen to be the
fluid four-velocity. This suggests that the quantities constructed in this way
are closely related to quantities in the Bardeen formalism in comoving
gauge. That is indeed the case, as we show below.

In the following we are using the definitions from
Ref.~\cite{Challinor:1998aa}, and only consider scalar quantities. We
choose here a particular quantity for comparison, though this is
without loss of generality. The ``comoving fractional density
gradient'' is defined in Ref.~\cite{Challinor:1998aa} as
\be
\label{defcalX}
{\cal X}_\mu\equiv\frac{a}{\rho}{}^{(3)}\nabla_\mu\rho\,,
\ee
and
\be
\label{defgradrho}
{}^{(3)}\nabla_\mu\rho\equiv h^\lambda_\mu\nabla_\lambda \rho\,,
\ee 
with the projection tensor 
\be
\label{project}
h_{\mu\nu}\equiv g_{\mu\nu}+u_\mu u_\nu\,,
\ee
and $u_\mu$ is a physically defined four-velocity which reduces in FRW
to the that of the ``fundamental observers'' ($\nabla_\mu$ is the
standard covariant derivative, which here however reduces to the
partial derivative because it is applied to a scalar quantity).\\

In order to relate the two formalisms, we now substitute the Bardeen
formalism (first order) quantities into the above. We have for the
components of the four-velocity $u_\mu$ (see e.g.~Ref.~\cite{MW2008})
and using conformal time $\eta$,
\bea
\label{def4vel}
u_\mu&=&a\big[ -(1+\phi),\ \p_i(v+B) \big]\,,\nonumber \\
u^\mu&=&a^{-1}\big[ (1-\phi),\ \p^{i} v\big]\,,
\eea
where 
is $B$ the shift function given in \eq{pert_metric}, and $v$ the scalar
velocity perturbation, and get for the density gradient
${}^{(3)}\nabla_\mu\rho$, to first order
\be
\label{covariant_delta_rho}
{}^{(3)}\nabla_i\rho=\p_i\delta\rho+\rho_0' \p_i \left(v+B\right)\,,
\ee
the temporal part being identically zero.
The right-hand-side of \eq{covariant_delta_rho} agrees, modulo
the gradient, with the expression given in Ref.~\cite{MW2008} for
the comoving density perturbation, namely
\be
\wt{\delta\rho}_\com\equiv\delta\rho+\rho_0'\left(v+B\right)\,,
\ee
which is what we expected.

\subsection{The issue of non-locality}
\label{nonlocal}

We finally reflect on another topic of frequent discussions with
colleagues, which was also discussed recently in
Refs.~\cite{Clarkson2011}, namely the issue of whether perturbation
theory is ``local'' or ``non-local'', with regard to metric and
covariant approaches.

The question here is, how to deal with gradient terms that appear in
the governing equations. Gradient terms are already present in the
definitions of the perturbed variables, e.g.~in the definition of the
4-velocity, \eq{def4vel} above, or in the spatial section of the
metric tensor, if we decompose the variables on spatial
3-hypersurfaces. This decomposition leads to the standard
``nomenclature'' for matter and metric perturbations, labelling them
by their transformation behaviour on the spatial 3-hypersurfaces: we
can decompose 3-vectors into a scalar (the curl-free part) and a
divergence-free vector, and similarly 3-tensors into scalars, vectors
(divergence-free), and tensors (divergence-free and traceless). This
decomposition is popular in cosmological perturbation theory, as at
linear order the governing equations for the different types of
perturbations decouple \cite{Bardeen80}, which simplifies the
calculations.

To get rid of the gradients in the ensuing equations, we have to
formally integrate the equations, or if we work in Fourier space,
solve the governing equations for the mode functions (see
e.g.~Refs.~\cite{MFB,Nakamura:2011hn}). This is straight forward at linear
order, where only simple operators of the form $\p_i$ (acting on the
whole equation), or $\p_i\p^i$ arise. 

However beyond linear order this is no longer the case, and to rewrite
the governing equations in closed form in terms of scalar quantities
we will inevitable pick up inverse gradient terms (see
e.g.~Refs.\cite{Bartolo_review,M2005,M2006}). It is these
inverse gradients terms that are usually referred to as ``non-local''.
This also means that in Fourier space this will lead to mode-mixing,
which requires solving convolution integrals.\\

There are several ways to deal with this issue:
%
a) We can use a large scale or super-horizon approximation and a
gradient expansion, that is simply neglect the gradient terms at a
particular order (see e.g.~Refs.~\cite{SB,Deruelle,M2005}).

b) Alternatively, we can keep the gradient terms as they are without
solving for the scalars, as is common in ``standard GR'' and the
covariant formalism, as advocated in
e.g.~Refs.~\cite{Ellis:1989jt,Clarkson2011} (at linear and higher
order, respectively). But then we can only calculate e.g.~the power
spectrum of the gradient of the density perturbation. To get the power
spectrum of the density perturbation itself, we will have to invert
the equations to get rid of the gradients.

We could also in the Bardeen formalism keep our equations ``local'',
if we do not invert the gradient terms. But then we would only have
governing equations for the gradients of scalars as in the covariant
formalism, and have the same problems as mentioned above.

c) Or we can simply suffer the consequences of deriving governing
equations for the scalar variables at higher order, namely the
introduction of non-local terms, and very complex equations. However,
although complex the ensuing equations can be readily solved
numerically (see e.g.~Refs.~\cite{M2006,Huston2009}).

To conclude this section: if we want governing equations for purely
scalar quantities, both the covariant and the Bardeen formalism will
lead to equations with inverse gradient or non-local terms.

\section{Conclusion}
\label{discussion}

As GR is a non-linear theory, we usually have to resort to some form
of approximative scheme to solve the governing equations. Perturbation
theory allows us to solve the equations iteratively order by order.

In cosmological perturbation theory we have to allow for perturbations
of spacetime itself, by using a perturbed metric tensor and
correspondingly perturbing the coordinates.
This is a major difference to classical perturbation theory in
e.g.~fluid dynamics, which uses a fixed Euclidean or Minkowsky
spacetime.\\

Gauge dependence in cosmological perturbation theory comes from the
need to work with a unique background spacetime, that remains fixed
under small coordinate transformations.
We can begin with the physical spacetime $\Me$ (see Fig.~\ref{fig1}),
on which we define all unperturbed variables. We then start our
perturbative expansion, by introducing a fiducial background spacetime
$\M0$ on which we define the background quantities, both matter and
metric variables.
The perturbations are then defined as the difference between the
quantities defined on the physical spacetime $\Me$ and the unperturbed
quantities on $\M0$.

We relate the background quantity to its perturbed version (on $\Me$),
i.e.~ get a correspondence between the perturbed and unperturbed
realisations of the quantities using a one-to-one map, parametrised by a
vector field, the gauge generator.
A gauge transformation changes this correspondence and hence the
perturbations unless they are gauge-invariant.  We can always
construct gauge-invariant variables, at any order, by studying their
transformation behaviour, and then combining the variables in such a
way that the gauge artefacts that they pick up in the transformation
cancel.
Note that a coordinate transformation transforms all the coordinates,
i.e.~in both the background and physical spacetimes, whereas a gauge
transformation does not affect the background quantities.\\

The gauge issue outlined in this note will arise in any theory that
keeps the background fixed under small coordinate transformations.
Once we have decided to use cosmological perturbation theory, a choice
of gauge is therefore always required if such a theory is used, even
when dealing with physically measurable quantities and working on
small scales. Hence dealing with the gauge issue is not optional or a
question of taste, it is dictated by the fact that we have chosen
cosmological perturbation theory and a unique background spacetime.
Of course, for many applications Newtonian theory will be sufficient,
and hence no gauge issue will arise.

Moreover, we should consider the gauge freedom as an opportunity to
choose the simplest form of the governing equations possible, not as a
problem. As long as we use the formalism with care, namely make sure
that no gauge artefacts or ``unphysical degrees of freedom'' are left
in the equations, there will be no ambiguity when we relate
observations with theory.

\begin{acknowledgements}
The authors are grateful to Ellie Nalson, Iain Brown, Adam
Christopherson, Juan Valiente-Kroon, and David Wands for useful
discussions and comments. KAM is supported in part by the STFC under
grants ST/G002150/1, ST/H002855/1, and ST/J001546/1.
\end{acknowledgements}

\appendix

\section{Covariance versus Gauge-invariance by example}
\label{example}

In this appendix we present a very simple example to highlight the
difference between \emph{covariance} or invariance to general
coordinate transformations, and \emph{gauge-invariance} as used in
cosmological perturbation theory, as defined in Section \ref{origin}.

To keep things simple, let us study  a scalar function dependent on a
single coordinate only, $\rho=\rho(t)$.
A scalar is by definition invariant under coordinate transformation
\be 
\label{def_scalar}
\rho(t)=\wt{\rho}\left(\tilde t\right)\,, 
\ee
where $\rho$ is the function evaluated in the coordinate system $x^\mu$
and $\wt{\rho}$, the function evaluated in coordinate system $\tilde
x^\mu$ (note that $t\equiv x^0$).

To see the effect of the different types of transformation, let's
assume we have the following simple dependence on $t$ for our
function,
\be
\label{def_func_rho}
\rho(t)\equiv t+ \epsilon t\,,
\ee
where for now we make no assumption about $\epsilon=const$, i.e.~we
here do not assume that $\epsilon$ is small.

We now introduce a new coordinate system $\tilde x$, related to the original
system $x$ by 
\be
\label{def_coord}
\tilde t=t+\delta t\,,
\ee
where again, we make no assumption about the size $\delta t$. From
\eq{def_scalar} and \ref{def_coord} we then get
\bea
\rho(t)&=&\rho(\tilde t-\delta t)\nonumber\\
&=&\tilde t-\delta t+ \epsilon\left(\tilde t-\delta t\right)
=\wt{\rho}\left(\tilde t\right)\,,
\eea
where the last line gives the functional dependence of $\wt\rho$ on
$\tilde t$.

\subsection{``Covariance at work''}

Let us first show the effect of a covariant transformation (no need to
properly perturb or expand things). We can now evaluate the function
$\rho$ at a particular point $Q$, with coordinate $t=\tc$, say, and find
that
\be
\rho(t=\tc) = \tc + \epsilon \tc\,,
\ee
rather unsurprisingly. The coordinate value in $\tilde x$ for Q is
$\tilde t= \tc +\delta t$, and substituting into $\wt{\rho}(\tilde t)$ we
get
\be
\wt{\rho}(\tilde t = \tc + \delta t)= \tc + \epsilon \ \tc\,,
\ee
that is the same functional value as in the system $x$, as required. \\

\subsection{``Gauge-dependence at work''}

To see the effect of the gauge transformation, we have to perturb and
expand our quantities, so let's assume $\epsilon\ll 1$ and $\delta t
=O(\epsilon)$. From \eq{def_func_rho} we can immediately read off the
background or zeroth order part and the perturbation of $\rho$,
namely
\be
\rho_0(t)=t\,, \qquad \delta\rho(t)=\epsilon t\,.
\ee
The crucial difference to the covariant case above is that now we require
for the background part of the variable, using \eq{Qback}, 
\be
\rho_0(t)=\wt\rho_0(t)\,,
\ee
that is the functional dependence of the background functions $\rho_0$
and $\wt\rho_0$ on the independent variable is the same (i.e.~in
cosmology we would want the same background FRW spacetime).

Evaluating our variables in the system $\tilde x$ we then have
\bea
\rho_0(t)=\wt\rho_0(\tilde t)=\tilde t\,,
\eea
for the background part, and
\bea
\delta\rho(t)&=&\epsilon t=\epsilon\left(\tilde t-\delta t\right)\nonumber\\
&=&\epsilon \tilde t + O(\epsilon^2)\,.
\eea
for the perturbation.

Again evaluating at the point $Q$, we have in $x$ that
$\rho(Q)= \tc (1+\epsilon)$, but in $\tilde x$ we have (to linear order)
$\wt\rho(Q)= \tc (1+\epsilon)+\delta t$, so there is discrepancy.

The standard result how the perturbation $\delta\rho$ changes under a
gauge transformation $t \to t+\delta t$ was given above in
\eq{gaugetrans_rho} (repeated here for convenience) 
\be
\wt{\delta\rho}={\delta\rho}-\dot\rho_0 \delta t\,,\nonumber
\ee
and we see that this fixes the discrepancy described above.



\end{document}